# Towards a Systems Engineering based Automotive Product Engineering Process

Hassan Hage[1, 2], Vahid Hashemi[2], Frank Mantwill[1]

[1] Helmut-Schmidt-University, Holstenhofweg 85, 22043 Hamburg, Germany
`hassan.hage@hsu-hh.de`
[2] AUDI AG, Auto-Union-Straße 1, 85057 Ingolstadt, Germany

**Abstract.** Deficit and redundancies in existing automotive product development hinder a systems engineering based development. In this paper we discuss a methodical procedure to eliminate deficits in the current product development and in turn to enable the introduction of a new systems engineering based development methodology. As the core part of our approach, we discuss how to transform an opaque heterogeneous product development to a homogenous consistent product development taking into account existing disciplines. Our approach paves the way to achieve a process structure that is more amenable to verification and validation. We show the effectiveness of our proposed solution approach on an automotive use case.

**Keywords:** Business Process, Systems Engineering, Verification & Validation.

## 1   Introduction

The ever-increasing demand for technology and connectivity in automotive industries has led automakers to invest heavily in the electronics and software development. The interaction of the three components hardware, electronics and software is becoming increasingly important, which in turn increases the complexity of a vehicle development [3, 20]. The complexity arises due to the fact that safe, environmental friendly, economical and easy accessible vehicle is demanded in the market [7]. At the same time, the legal requirements, which are required for an initial vehicle registration, are tightened [15]. Despite increasing complexity in a vehicle, a reduction in development times due to competition is necessary [1, 20]. Hence, automotive manufacturers are facing an exponentially growing challenge for which a reformation of their development strategy is required [20, 25]. To account for the latter, manufacturers orient themselves according to the standard of the "Automotive Software Process Improvement and Capability Determination" (A-SPICE), which should enable a mastery of the development complexity [8]. Hence, the development methodology of Systems Engineering (SE) plays an important role, since this methodology will help to master the complexity. Systems engineering approach has been exploited for many years in avionics industries for product development. The increasing product complexity in automotive industries solicits application of this method [4] which in turn is associated with complications.

Firstly, vehicle manufacturers have grown dramatically since they were founded in the past few decades. Due to the drastic and rapid expansion of the company, a partly heterogeneous company structure was formed. Domain-specific departments independently developed their own processes, methods and tools [13]. As a result, it was no longer possible to ensure a homogeneous, consistent and transparent product development. Therefore, the traceability in the product development is sometimes challenging and hence, the high product quality can be only ensured with difficulty. The recent vehicle software manipulation for incorrect exhaust values is an example of a product development in which such manipulations were difficult and time consuming to be traced [7, 11]. The reason for such a difficultly is the heterogeneous processes of the product development.

Moreover, in the automotive industry, the Product Engineering Process (PEP) traditionally consists mainly of individual phases and control points/milestones. Behind the phases are processes that have to be reached at certain milestones [12, 24]. Generally a PEP is saved in a company as a specific document format, which is available for retrieval. However, the processes presented in the PEP do not provide the domain-specific departments with process-related, sufficiently detailed information. In particular, the necessary detailed process steps to reach the milestones are not apparent. Consequently, there exists no product development process overview which could deliver sufficient information and the interdependency of all relevant groups of a product development. In addition, the scheduling for domain-specific departments is very rough and dependencies between these departments and intermediate milestones are not clear. Finally, the use of the SE method requires a radical change in the corporate culture as well as parts of the corporate roles, processes, methods and tools as the current product engineering process is not designed based on the SE.

The present paper presents a methodological approach that arose out of necessity during the attempt to introduce SE at an automotive manufacture and resolves current deficits in existing automotive developments that make it difficult to introduce new development methods such as the SE.

Summarizing, the main contributions of this paper are as follows.
- We introduce and integrate milestones in process models to capture temporal aspects of the product development.
- We show that by means of a unified modelling language in a heterogeneous product development, the connectivity from the highest level of abstraction down to the lowest level of abstraction can be ensured. This way we provide consistency and traceability in the entire product development.
- We provide an approach which enables SE in the automotive product development. We show promising results on the feasibility of our solution approach, obtained by its implementation on an automotive use case.

**Structure of the paper.** Section 2 provides a brief overview of the related works and highlights the research gap. In Section 3, a solution approach is described. This approach promises the elimination of deficits of an automotive product development and presents a method to enable the introduction of SE in an existing product development. We then validate the feasibility of our proposed approach on a use case. Finally, Section 5 discusses the solution approach and concludes the paper.

## 2      Related Work

This section provides an overview of related works. Existing methodologies, which should provide transparency and consistency in the product development, are briefly examined for advantages and disadvantages. This leads us to identify the current gap and therefore, to propose a solution approach.

### 2.1      Methodology for Product Planning

The organization of German engineers a.k.a. "*Verein der Deutschen Ingenieure (VDI)"* proposes a methodical procedure according to VDI 2220 guideline to plan the life cycle of a product [22, 23]. This life cycle is referred to as a product planning or in a corporate level as a product process (PP) [12]. It starts with a market research through the PEP and finally, up to the product disposal [9, 22]. The actual product development takes place in the PEP and includes the planning, drafting, designing and realization of a new product. To account for the latter, the VDI guideline presents a methodical approach on a detailed level. Finally, following the guideline leads to a consistent and traceable product process. Product developments of automotive manufactures are roughly similar (see Fig. 1).

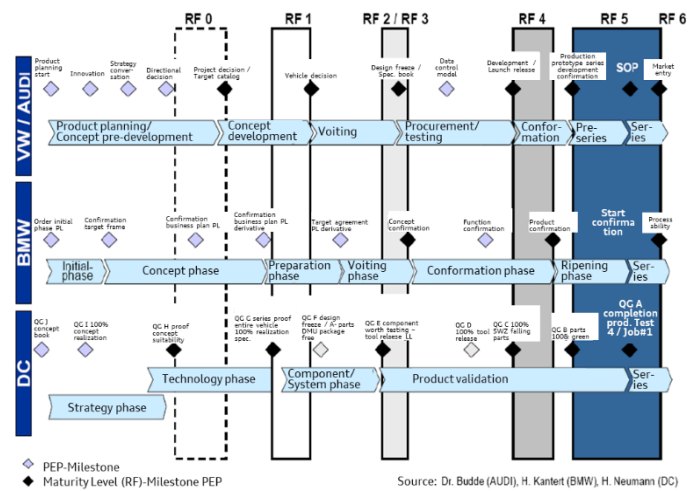

**Fig. 1**. Product Engineering Process of various automotive manufactures [19].

As mentioned in section 1, a product process as well as a PEP consists of phases and milestones. This is a crucial bottleneck as product development according to VDI 2220 does not consider the temporal aspect of a development. Moreover, the guideline VDI 2220 specifies a strict method; however, in practice a development consists of huge amount of methods. Therefore, following such a strict method is hardly realizable. After all, the VDI 2220 guideline is not designed for complex mechatronic products [22, 23].

**2.2     Design Methodology for Mechatronic Systems**

The organization of German engineers introduced the guideline VDI 2206 for designing of mechatronic systems [23]. This guideline consists of elements problem-solving cycle as a micro-cycle, the V model as a macro-cycle as well as process modules for recurrent working steps. The first problem-solving element is based on the method of SE but further considers disciplines such as business administration which are necessary for a product development. Target of this element is to analyze the initial problem as well as to determine the actual and desired state. The second element consists a problem solution and plans the further procedure based on the V model of SE for developing mechatronic systems [23]. Finally, the third element defines and deepens a part of the V model with the use of process modules. This approach enables a consistent and traceable development of complex mechatronic products by strictly following the guideline. Nevertheless, this approach similarly to the VDI 2220 guideline enables a theoretical approach based on a completely new product development. Therefore, already established processes, methods and tools could not further be used and an existing development has completely to be changed. The latter is unrealizable due mainly to the high product complexity as well as the lack of consideration of temporal aspects.

**2.3     Systems Engineering / Model-Based Systems Engineering (MBSE)**

Systems Engineering is an interdisciplinary approach for the development of multidisciplinary systems such as mechatronic systems. With the help of this methodology, a holistic and cooperative understanding between all development participants is created [15, 21]. However, it does not consider necessary disciplines such as business administration and temporal aspects for the product development. The SE approach is oriented on the V model. The V model is started by the RFLP approach – R stands for Requirements, F for Functional Model, L for Logical Model and P for Physical Model – through hardware, electronics and software development up to verification and validation of the entire product [17, 18]. A V model-based vehicle developments is shown in Fig. 2. This development process requires specific processes that are performed with specific methods and tools [18]. MBSE is a method of SE and aims to achieve a largely model-based development through the V model. For example, dependencies between functions are recorded as a model and not as a text [13]. MBSE offers the advantage of a uniform standard language amongst the product development so that consistency and traceability are easier to realize. Moreover, MBSE creates a better holistic understanding between developers [13]. Nevertheless, using MBSE over the entire existing product development needs a radical change of processes, methods and tools. In particular, use of this approach is not realizable when the existing development is not transparent as detailed in section 1.

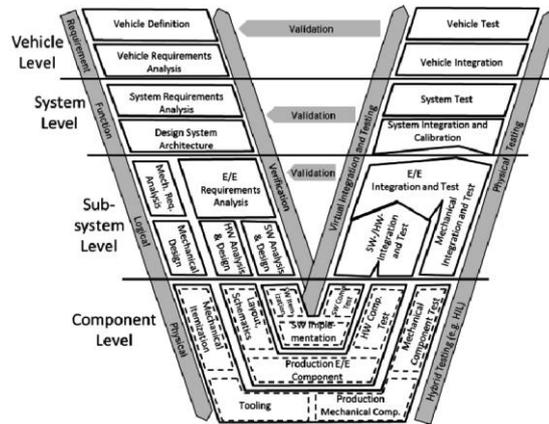

**Fig. 2:** V-Model of SE for developing a vehicle [17].

### 2.4   Research Gap

Following strictly the aforementioned approaches enables a consistent and traceable product development. The main issue is that all approaches are based on a new product development including specific processes, method and tools. However, a method for reaching consistency and traceability in an existing development is required. Additionally, necessary disciplines such as production, marketing, finance, temporal aspects etc. shall be considered in a product development. To account for the latter, a method shall be developed to enable transformation of an opaque heterogonous product development to a homogenous consistent product development taking into account existing disciplines as well as processes, methods and tools. The latter is clearly identified as a research gap which is the core focus in this paper. It is worthwhile to note that the required method has to inevitably consider a development based on SE as automotive manufactures orient themselves according to A-SPICE guidelines and the approach of SE. However the integration of the SE approach shall be incremental as already developed processes, methods and tools are not possible to change abruptly.

## 3   Solution Approach

In order to record the actual state of the product development in terms of processes, the top-down approach "from rough to detail" is used. This approach recommends breaking down an overall problem to be solved into logical and interrelated sub-problems. The sub-problems are further broken down until small manageable problems are obtained which can then be solved [2]. The bottom-up approach could also used but for the target of this paper the bottom-up approach is more time intensive. The recording of the actual state is done in a similar way as the top-down approach. At the beginning the business process is defined. Subsequently, the specification of the main process takes place. After that associated sub-processes are specified until the last process workflow level is finally recorded. Fig. 3 reflects the process hierarchy where *X* means the last specified sub-process and *Y* the last specified process workflow.

In a complex overall process, sub-processes are designed by different people with different understandings and background knowledge. For this reason, an existing specification language is used to ensure uniform understanding. The Business Process Model and Notation (BPMN) language [10] is recommended for this purpose because this language enables a sufficiently described process through its notations. All recorded processes are represented using the BPMN specification language. It is also possible to create logical dependencies between the main- and sub-processes. In practice, the success of a development does not only depend on the process flow but also on the adherence to the time component and milestones. For this reason, it is necessary to take the time component into account at the process specification and modeling phase. Specification languages like BPMN do not offer suitable symbols to capture milestones in the process model. Therefore, we provide symbols from the BPMN language with a different context in this approach.

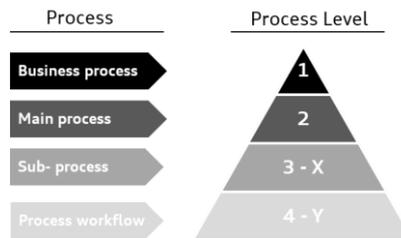

**Fig. 3.** Process Levels.

The top-down process modeling and specification are prerequisite for enabling the inevitable usage of SE. However, as already discussed in Section 2, the application is not straightforward for an existing automotive development. To this end, and with the aim of enabling the SE approach, we discuss a methical approach to transform an existing product development to achieve consistency and, in the case of e.g. changes, the possibility of traceability. The high level idea of our approach takes the prepared process modeling and specification and compares that with the prescribed processes of the SE to observe similarities and deviations. The procedural changes cannot be avoided during the implementation of the SE as the nature of SE is up to a certain extent, theory-based. Therefore, it is necessary to obtain a constant overview of the quality of the changed processes in order to remain within the scope of the prescribed SE process standard. Our approach includes implementation of five phases detailed as follows.

### 3.1 Phase 1: *SP – Setting a Pyramid*

A product is described as an entire system which in turn consists of several subsystems, each of which are composed in components. The development of such systems or components requires a large number of specialist, who are simply responsible for the system or component. In order to achieve a consistent product development, it is necessary to make the existing structure transparent, such that transparency is realized through the cycle shown in Fig. 4.

The starting step *Determine Level* identifies a sub-process of the product process and determines its hierarchical assignment (see Fig. 3) to the entire process. Then, the next step *Determine Docs* captures all required documents that are needed to perform the associated process. The focus here is especially on process flow overview including timing dependencies. Finally, the last step *Adding to Pyramid* takes all processes including documents to the associated level in a notional pyramid down.

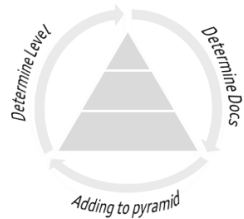

**Fig. 4.** Cycle of Process Transparency.

### 3.2 Phase 2: *NL – Neutralizing the language*

Complex product developments usually involves a large number of different, independent processes, methods and tools. For example, Microsoft Excel (ME) is used in a certain sub-process. Therefore, results are generated and forwarded in an ME format. On the other side, an in-house developed tool is used to generate results in another sub-process. The large number of sub-processes and their complexities result in a difficult communication between different processes. Therefore, a lot of time is invested in translating technical terms from other domain-specific departments [7]. Besides, it is not possible to assign a domain-specific process to the entire process chain or product development. As an immediate consequence, there are often unwanted, redundant sub-processes. To avoid these problems, a common holistic modeling language is required, which is performed in this phase. All processes recorded in the first phase has to be translated in a holistic language. This requires a graphical specification language as process design is done by various people with different understandings and background knowledge. Therefore, a specification language accessible to all process-involved persons is necessary. As mentioned earlier, the solution approach of this paper recommends the BPMN language for process modeling.

In order to obtain an expressive enough process model, some qualitative requirements has to be taken into account to ensure that all for a consistent product engineering process required information are for the further procedure guaranteed:

1. Only necessary process steps are recorded.
2. Each process step has a defined *input* and *output*, and a specified *execution time*.
3. All *actors* within a process are assigned to a specific *role* description.
4. Milestones are in the process model equal to *events* [10]. A start milestone is equal to a *start event*, an intermediate milestone (see Fig. 5.b) is equal to an *intermediate event* and an end milestone is equal to an *end event*. For covering the temporal aspect, the event symbol is always followed by a time symbol, depicted in Fig. 5.a. This symbol describes the time that is needed to reach an event. In total, the time and event symbol cover a milestone.

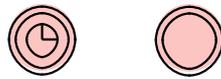

(5.a) time symbol    (5.b) intermediate event

**Fig. 5.** The symbols depicted here are captured by the Software *Enterprise Architect*.

These qualitative requirements must be observed when creating process models. All of the processes captured in phase 1 are continuously modeled according to these requirements.

### 3.3   Phase 3: *LD – Logical Dependency*

This phase deals with the development of a logical process structure to achieve the entire consistent process. For this purpose, all relations between processes with each other are encoded by means of the specification language. Fig. 6 shows a quantitative representation of logical dependencies.

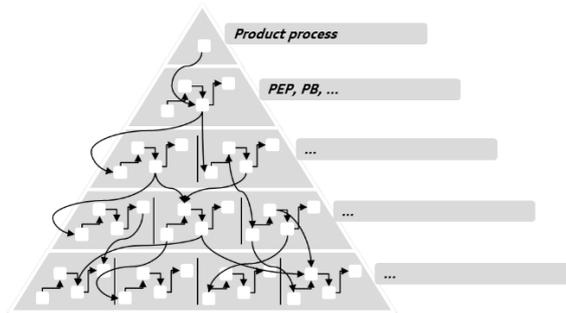

**Fig. 6.** An entire logical and coherent process.

In general, the lowest level of the pyramid has the highest level of detail, since this is the place of the domain-specific development. In contrast, the top pyramid level has the highest level of abstraction. Following the top-down approach, the highest level of abstraction is used to create the logical process structure. Starting from the uppermost process of the pyramid level, this process has to be connected to the process of the subsequent pyramid level. Due to a uniform specification language, a connection between the processes is possible without any complications. Afterwards, the second pyramid level is connected to the third level. This process is carried out from level to level until the last pyramid level is reached. An entire continuous process is achieved when the dependency of a process on all other processes at all levels can be shown. Dependencies among processes at the same pyramid level are given at the next higher level and denoted as implicit dependencies. Section 3.4 deepened the way to reach these dependencies.

### 3.4 Phase 4: *CBM – Controlled by Milestones*

The phase *Controlled by Milestones* aims to detect implicit dependencies of processes and provides all target groups of the pyramid with all common process relevant information. This is realized by symbols of a graphical specification language. The time symbol shown in Fig. 5.a describes a specific time or date that supports an intermediate process to trigger or completes a process. Each event that occurs between the beginning and the end of an event is called an intermediate event as depicted in Fig. 5.b.

An event is equated with a milestone in the present context. In other words, a milestone in a process model is reached when an event is reached after the process steps have been successfully completed. Since a milestone always has a temporal dependency, a time symbol is added to the intermediate event symbol in order to transfer the meaningfulness of a milestone completely in a process model.

As mentioned before, milestones are represented in the process model with the help of the time and event symbols. The milestone *Project Start (PS)* is equal to an intermediate event and the time specification fictionally indicates that the milestone PS occurs at the specified time before *Start Of Production (SOP)*. It is worthwhile to mention that the idea of covering temporal aspects is partly considered in the work of [6]. This work describes the usage of time symbols for every process step in a BPMN model. Each process step/task obtains a time symbol that describes the time needed for performing the step/task. Nevertheless, this approach does not take into account the idea of milestones into a process model such as the BPMN model. The latter has been addressed in our approach through simultaneous use of time and event symbols.

An example for introducing a milestone into a process model is depicted in Fig. 7. It is quite often that milestones are represented in the shape of inverted triangles with a given name and time.

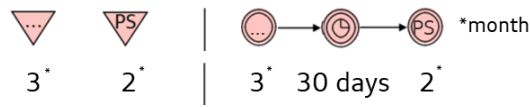

**Fig. 7.** Capturing milestones in BPMN. In this example, a previous time event occurs for three months before SOP and one month passes until the intermediate event PS is reached. The intermediate event PS thus occurs two months before SOP.

Every process model has a process flow which is performed by defined roles with the help of one or more tools at a given time.

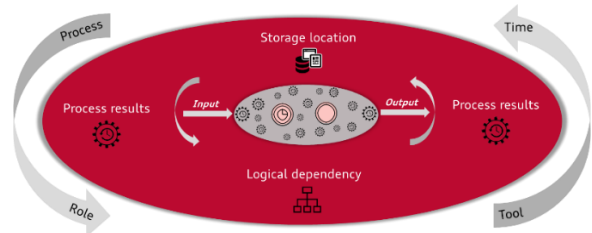

**Fig. 8.** Milestone life cycle.

In order to start a process, at least one input variable is required. The process in turn provides at least one output variable. A milestone life cycle is depicted in Fig. 8. It describes all necessary information which a milestone should have in order to detect implicit dependencies and to realize a consistency. A milestone requires answering of the eight *Golden Questions (GQ)* for obtaining all relevant information to build up a consistent and traceable BPMN model:

- GQ1 - Which process has to be performed?
- GQ2 - Which role is responsible for the process?
- GQ3 - Which tools are required?
- GQ4 - How much time is necessary for performing the process?
- GQ5 - Which input is required?
- GQ6 - Which output is generated?
- GQ7 - For which milestone is the generated output of GQ6 required as an input?
- GQ8 - On which file are the inputs/outputs saved?

If the eight GQs are answered for each milestone of the process, the processes at various levels are able to know interdependencies through the milestones. The answers of the eight GQs have to be linked to each milestone. According to the BPMN standard, inputs and outputs could be described in a process model with the symbol Data Object.

### 3.5 Phase 5: *CSEA – Compliance with Systems Engineering Approach*

This phase deals with the qualitative changes of processes, methods and tools. The processes, methods and tools that have been established for years need to be revised or reformed for introducing the methodology of Systems Engineering. This is done based on the automotive V model of SE as shown in Fig. 2.

The linked processes from the pyramid of phase 4 are compared with the processes of the V model for similarities and deviations. Four aspects are considered in this comparison; namely, *Process steps/tasks, Roles, Methods and Tools.*

The processes at the lowest level of the pyramid need to be compared with similar processes of the V model. For example, in an automotive development components of a vehicle are tested before approval. The V model also provides a component test. All necessary process steps in the existing process for testing a component has to be compared with the process steps for component test according to SE. If process steps differ, a revision of the old process according to the V model process is required. It should be revised for the entire process otherwise a consistent product engineering process according to SE is not ensured. This applies also to the roles, methods and tools. This comparison including changes is elaborated for each individual process. It is possible that old processes deviate strongly from processes of SE but then new process has to be added and old one removed.

The V model of SE is characterized by verification and validation of developed systems and functions. For ensuring that characteristic, processes at the right side of the V model has to be linked to the processes of the left side of the model. Consequently, roles that are responsible for testing obtain the possibility of verification and validation by calling up e.g. the target requirements. Moreover, during revising the product engineering process sufficient iterations for verification and validation of different degrees of maturity have to be considered. In addition, in further procedure methods and tools of the corresponding processes could be used to stabilize verification and validation. This requires a comprehensive networking in the entire product engineering process of all tools which is out of the focus of this paper. It is worthwhile to note that in order to ensure the possibility for verification and validation during the implementation of the SE processes the retaining of the entire milestones in the model shall be considered. The latter is due to the fact that exchange of the performance and content of the processes are triggered by means of milestones. This guarantees verification and validation of the process at a proper time with the corresponding process content. Iterations of verification and validation can be increased arbitrarily by adding intermediate milestones in the process model. Through a successful implementation of SE processes a traceability on each process level of abstraction can be ensured through the entire process. This in turn enables recognizing process changes and obstacles in time.

## 4     An Automotive Use Case

The feasibility of the proposed approach in Section 3 has been tested to a certain extent on an automotive use case. Since such a check is not feasible for the entire vehicle development within the framework of a research project, we chose *testing process of a park pilot* as a use case. The aim is to integrate and demonstrate this use case in the context of the entire product development and to design its SE compatible process. Initially, phases 1 and 2 of the solution approach are applied. Thereby all levels of product development, which are relevant for the use case could be determined. Following the top-down approach, the Product Process is linked through the product engineering process to the Function Chart which in turn is linked through a Test Plan down to the lowest level where the required department was located as depicted in Fig. 9.

In parallel, all interdependent processes, methods and tools are recorded. Information about the processes at different levels are obtained from various stakeholder with different data formats. After determining all necessary processes and their contents, phase 3 to 5 are applied, so that finally a uniform overall BPMN model is created that starts from the Product Process. From the Product Process all levels can be broken down and navigated until finally the process of the use case testing of a park pilot is reached on the lowest level.

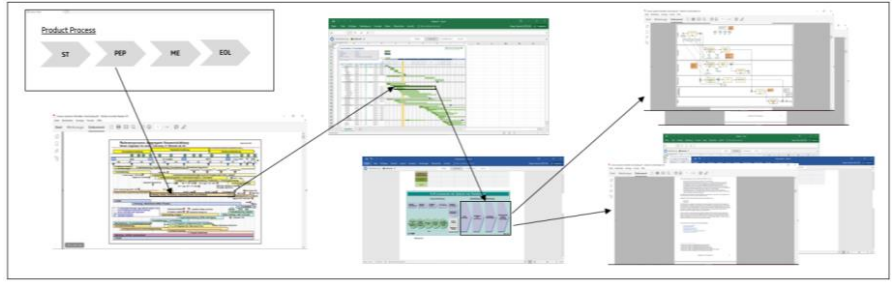

**Fig. 9.** Current Product Process.

The corresponding BPMN model in a coordinate system is described in Fig. 10. The overall process of vehicle development, the Product Process, begins at the origin. To reach the process of the testing of a park pilot, a constant deepening of the processes is necessary. At the same time the complexity of the processes increases due to the technical details. Finally, the importance and dependence of the testing of a park pilot use case in the overall context could be determined. More precisely, if this process is changed the effect of this change on the overall process can be tracked.

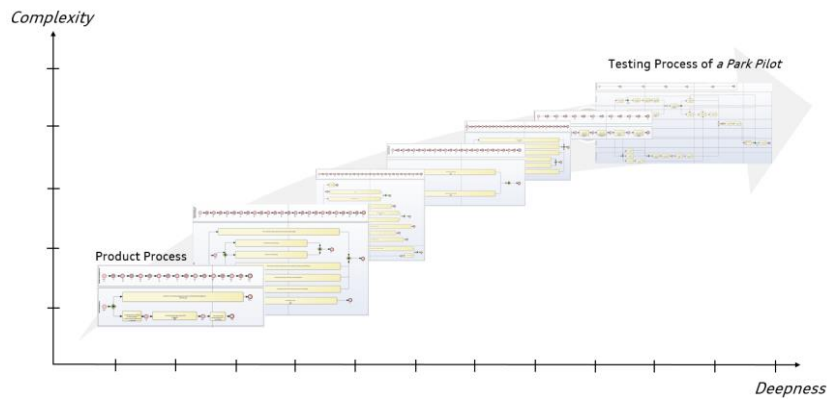

**Fig. 10.** BPMN model of the entire process in a coordinate system.

Each BPMN diagram in the model is given a timeline which can be seen as milestones, as depicted in Fig. 11. This figure describes the timeline of the entire product process at a high level with its associated milestones. Hence the time aspect in the process could be taken into account. Furthermore, the timeline helps to enable dependencies between the processes by means of input and output variables. The differences between the existing process of the testing of a park pilot use case and the processes to be achieved according to SE are determined in the last phase. For this use case we analyzed the testing process of a park pilot and could notice only small deviations between the existing process and the process according to SE. According to the deviations the old process has been revised to be SE compatible.

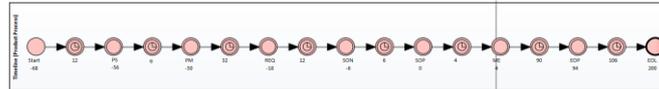

**Fig. 11.** Timeline of the Product Process.

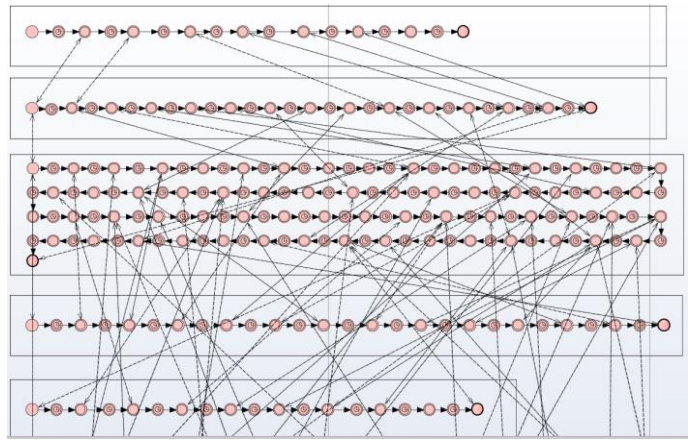

**Fig. 12.** Excerpt from the networked milestones at different process level.

In this use case only small process modifications were required. However, this is not always the case. Integration of different SE processes may require massive modifications which in turn are time consuming. A further investigation may be needed to automate adaptation of the old processes to the SE ones. However, such an automation is not in the focus of the present paper. It is very important to follow the structure of revising processes step by step. Otherwise, the idea of verification and validation by means of milestones is hindered. Therefore, during the implementation of SE milestones correct interdependencies are needed as depicted in Fig. 12. In this figure, a cut-out of the interdependencies of the milestones at several levels of abstraction of the entire product is presented. The third rectangle in Fig. 12 depicts a reference timeline of milestones in one month step for the entire product process. This reference timeline ensures the correct link between milestones of higher and lower levels of abstractions. Moreover the link between different levels of timelines supports the transparency about inputs/outputs between various processes.

## 5      Conclusion and future work

Due to the increase in product complexity, the introduction of the SE methodology for vehicle manufacturers is inevitable as this methodology is necessary for achieving a high A-SPICE standard where the strong focus is on process qualities. In this paper, we identify deficits in the development process at vehicle manufacturers that impair process quality and develop a methodical approach to address this issue. Our proposed solution turns existing processes into a consistent structure and thus increases the existing process quality. At the same time, it enables adaptation of the SE approach to change development methodologies of a decade-old product development. Several product processes in the automotive industry are simultaneously existing for a certain number of products. In this regard, our solution approach enables exchange between different product processes through the phases discussed in section 3, especially the use of a uniform modelling language. Our proposed, unlike conventional methodologies, offers direct practical relevance. The effectiveness of the solution approach has successfully been tested on an automotive use case.

The SE is characterized by iterative verifications and validations on different development levels (e.g. components or subsystems). In order to make this possible, an extension of phase 5 is needed to be provided as the industrial PEP does not further consider the verification and validation methodology of the SE. Moreover, the identified processes shall be checked for their process-related and temporal implementation in practice. For this purpose, a logic is required that can record individual process steps, including the executed time of a developer and assign them in the entire process. This way deviations between the prescribed process and the executed processes in terms of content and time are to be detected and eliminated at an early stage. Furthermore, this logic should contribute to the optimization of the processes. Finally, an automated verification and validation shall be carried out which, at the same time, is compatible with SE.

**Acknowledgments.** This work is supported by the Helmut-Schmidt-University in Hamburg and by the AVAI project at AUDI AG in Ingolstadt.

**References**

1. Bögemann, Ingo. Product life cycles are getting shorter, your development times too? MB Collaborations , 2018.
2. Brugger, Ralph. IT-Projekte strukturiert realisieren. Springer Fachmedien Wiesbaden GmbH, 2003.
3. Brünglinghaus, Christiane. Elektronik und Software beherrschen Innovationen im Auto. Springer Professional, 2014.
4. Busch, Arno. Automotive SPICE: Die "Gewürzmischung" für System- und Softwareentwicklung. Continental Automotive GmbH, 2020.
5. Düchting, Carsten. Aufbau eines freigabe- und kommunikationsbasierten Assistenzsystems im Produktentstehungsprozess. University of Dortmund, 2005.
6. Duran, Francisco, Camilo Rocha, and Gwen Salaün. Stochastic Analysis of BPMN with Time in Rewriting Logic. 2018.

7. DW Made for minds. www.dw.com. 03 20, 2018. https://www.dw.com/en/bmw-searched-over-suspicious-emissions-software/a-43055629.
8. e.V., Verband der Automobilindustrie. www.vda-qmc.de. 2020.
9. Eßmann, Volker. Planung potentialgerechter Produkte. Springer Fachmedien Wiesbaden GmbH, 2013.
10. Group, Object Managment. www.omg.org. 2020. https://www.omg.org/spec/BPMN.
11. Hans-Dieter Zollondz, Michael Ketting, Raimund Pfundtner. Lexikon Qualitätsmanagement: Handbuch des Modernen Managements auf der Basis des Qualitätsmanagements. De Gruyter Oldenbourg, 2016.
12. Hans-Hermann Braess, et al. "Produktentstehungsprozess." Springer Fachmedien, 2013.
13. Hart, Laura E. Introduction To Model-Based System Engingeering (MBSE) and SysML. 2015.
14. Hutterer, Philipp. Reflexive Dialoge und Denkbausteine für die methodische Produktentwicklung. Technical University Munich, 2005.
15. Jürgen Gausemeier, et al. "Studie: Systems Engineering in der industriellen Praxis." Carl Hanser Verlag GmbH & Co. KG, 2013.
16. Klöckner, Jürgen. "Autohersteller kämpfen gegen EU." WirtschaftsWoche, 2013.
17. Marco W. Groll, Dominik Heber. "E/E-Product Data Managment in Consideration of Model-Based Systems Engineering." IOS Press, 2016.
18. Ralph Stelzer, et al. EEE Methoden und Werkzeuge in der Produktentwicklung. TUDpress Verlag der Wissenschaft GmbH, n.d.
19. Reuter, Martin. Technischer und wirtschaftlicher Vergleich von Herstellungsverfahren bei der Entwicklung von Kunststoffhohlkörpern in Automobilanwendungen. 2013.
20. Schömann, Sebastian O. Produktentwicklung in der Automobilindustrie. Springer Gabler, 2012.
21. Shortell, Thomas M. INCOSE Systems Engineering: A Guide for System Life Cycle Processes and Activities. John Wiley & Sons, 2015.
22. VDI Fachbereich Produktentwicklung und Mechatronik. "Systematic approach to the development and design of technical systems and products." Beuth Verlag GmbH, 1993.
23. Verein Deutscher Ingenieure. "Design methodology for mechatronic systems." Beuth Verlag GmbH, 2004.
24. Walla, Waldemar. Standard- und Modulbasierte digitale Rohbauprozesskette. KIT Scientific Publishing , 2017.
25. Watanabe, Katsuaki. "Toyota Motor Corporation." 2007.